\documentclass[prl,showpacs,amsmath,amssymb,aps,floats]{revtex4}
\usepackage{epsfig}
\usepackage{dcolumn}
\usepackage{bm}

\def\nnd{\end{document}}

\def\nnb{\nonumber}

\def\be{\begin{equation}}
\def\ee{\end{equation}}

\newcommand{\bea}{\begin{eqnarray}}
\newcommand{\eea}{\end{eqnarray}}
\newcommand{\bwt}{\begin{widetext}}
\newcommand{\ewt}{\end{widetext}}

\def\eed{\end{document}}

\def\al{\alpha}

\def\m_z{m_{\textrm {Z}}}
\def\al{\alpha}
\def\be{\beta}

\def\al{{\alpha}}

\def\mbkf#1{\big [ #1 \big ]}

\def\sBkf#1{\Big ( #1 \Big )}

\def\sBBkf#1{\Bigg ( #1 \Bigg )}

\def\Journal#1#2#3#4{{\it #1} {\bf #2}, #3 (#4)}

\makeatother

\begin{document}

\title{Spectra, triple and quartic gauge couplings in a Higgsless model}
\author{Kingman Cheung$^{1,2}$}
\author{Xiao-Hong Wu$^3$}
\author{Qi-Shu Yan$^{1}$}
\affiliation{
$^1$ Department of Physics, National Tsing Hua University,
Hsinchu, Taiwan \\
$^2$
Physics Division, National Center for Theoretical Sciences, Hsinchu, Taiwan\\
$^3$ School of Physics, Korea Institute for Advanced Study,
         207-43, Cheongryangri 2-dong, Dongdaemun-gu, Seoul 130-722, Korea
}

\begin{abstract}
Spectra, triple and quartic gauge couplings of the Higgsless model with gauge 
group $SU(2)_L\times SU(2)_R\times U(1)_{B-L}$ defined in warped space
are explored with numerical method. We extend the equation of motions, boundary conditions
and formalism of multi-gauge-boson vertices to the Hirn-Sanz scenario. 
By assuming the ideally delocalized fermion profile, we study 
the spectra of vector bosons as well as the triple and quartic gauge couplings among vector bosons. 
It is found that mass spectra can be greatly modified by the parameters of 
QCD power corrections.
Meanwhile, the triple and quartic gauge couplings can deviate from 
the values of the standard model to at least $\pm 10\%$ and can saturate the LEP2 bounds. 
We find the triple gauge couplings of $Z{\bar W}W$ 
can be $ 50 \%$ smaller than the unitarity bounds.
While the triple gauge couplings of ${\bar Z}WW$ 
is $20 \%$ smaller than the unitarity bounds, which might
challenge the detection of ${\bar Z}$ via s channel at LHC if $m_{\bar Z}> 500 \,\textrm{GeV}$.
\end{abstract}

\pacs{11.15.Ex, 11.30.Rd, 12.60.-i, 12.60.Cn}

\maketitle

\renewcommand{\thefootnote}{\arabic{footnote}}
\setcounter{footnote}{0}

\section{Introduction}
Electroweak symmetry breaking mechanism is one of the most important questions 
of the standard model (SM) which will be investigated by the LHC soon. 
Following its low energy QCD counterpart, and when only massless photon and massive weak bosons
are considered, the electroweak symmetry breaking sector can be effectively described by the electroweak 
chiral Lagrangian (EWCL) \cite{Appelquist:1980vg,Appelquist:1980ae,Longhitano:1980iz,Longhitano:1980tm,
Appelquist:1993ka} in a model-independent way.  Recently, it was proposed
by \cite{Dutta:2007st} to make a global analysis on the electroweak symmetry breaking sector 
by incorporating triple gauge coupling measurements. It was found
that the two-point chiral coefficients $\al_0$, $\al_1$, and $\al_8$ are 
tightly constrained from Z-pole data as well as W boson mass and top quark mass 
measurements. However, the triple and quartic gauge couplings are loosely bounded,
which can make the QCD-like models to be consistent with electroweak precision data.
At the same time, driven by the 
measurement of three-point and four point chiral coefficients at LHC and future 
colliders \cite{forwardtagging1,forwardtagging2,forwardtagging3,Boos:1997gw,Han:1997ht,Haywood:1999qg,Beyer:2006hx}, 
it is necessary to extend the theoretical study from two-point vertices to three-point and four-point vertices of the
EWCL. Therefore it will be interesting to look at those strongly coupled models where might have 
large deviations in the triple and quartic gauge couplings.

It is of great importance to measure the triple gauge couplings 
at LEP2 and Tevatron \cite{Hagiwara:1986vm, Ellison:1998uy}, 
because such a measurement is the 
direct proof of the non-Abelian structure of the standard model.
It can help to measure the size of $W$ and $Z$ weak bosons, and 
to understand to what extent these weak bosons can be interpreted as
the composite or elementary objects. Meanwhile, it can remove the theoretical
background for the discovery of new physics, for example to remove the SM background (say,
Z decaying to a WW pair) for the Higgs decaying to a WW pair.

Up to now, both LEP2 and Tevatron have measured the triple gauge couplings by
observing diboson events. D0 and CDF collaborations
from Tevatron have released their results on the single W production events \cite{Acosta:2004it}, 
WW pair production events \cite{Abazov:2004kc} and WZ pair production events \cite{Abazov:2005ys}. 
But the most stringent bounds on the triple gauge couplings are from LEP2 W pair production events. 
Till now, LEP2 already accumulate more than $40,000$ W pair events \cite{Axs,Dxs,Lxs,Lp,Op,Dp,Lc,Azzurri:2006na}. 
Total cross section, angular distributions and spin density matrix have been analyzed. 
The component of $\rho_{00}$ of the spin density matrix clearly show the existence
of the longitudinal component of the W bosons. Currently, the most stringent bounds
on triple gauge couplings are extracted from LEP2 data \cite{Heister:2001qt}. The bounds from
LEP2 \cite{Axs,Dxs,Lxs,Lp,Op,Dp,Lc,Azzurri:2006na, Dutta:2007st} on 
three-point chiral coefficients are a few $10^{-2} \pm 10^{-2} $ in one-parameter fit and
$10^{-1} \pm 10^{-1} $ in two-parameter fit.

It is quite natural to ask whether there exists a concrete example in strongly interacting
model where three-point functions can have large deviations from the predictions of
the SM. The answer provided in this paper is affirmative, we explicitly show that the deviation
can be as large as $10\%$.

It is well-known that it is extremely difficult to tackle strongly interacting dynamics.
Currently the only reliable tool is the lattice calculation.
Computing two-point functions with lattice method is relatively easy. However, computing
three point and four point functions is expensive and difficult.

Recently, the development in AdS/CFT correspondence \cite{AdSCFT}
provides a powerful tool to tackle strongly interacting dynamics. 
By assuming the AdS/CFT correspondence, 
it is possible to explore the strongly interacting gauge theories by studying 
the weakly interacting gravity theories. Spontaneous symmetry breaking can be 
realized in the gravity theories by imposing the boundary conditions. For example, 
the Higgsless model \cite{Csaki:2003dt, Agashe:2003zs, Csaki:2003zu}, 
the electroweak symmetry breaking is realized with boundary conditions imposed 
on the ultraviolet (UV) and infrared (IR) branes.
These boundary conditions can be interpreted as
composite Higgses developing vacuum expectation values on the boundary branes.
Such a model can be viewed as the AdS dual of a walking technicolor-like model.

Higgsless model, as its name indicated, has no Higgs boson as the particle content.
In the SM, the presence of the Higgs boson (lighter than 1 TeV or so) is necessary
to guarantee the unitarity of the scattering amplitudes of longitudinal 
gauge bosons to any energy scale above TeV \cite{Cornwall:1973tb}.
The Higgsless models suggest that the absence of Higgs is still possible and workable.
The unitarity of longitudinal gauge boson scattering amplitudes 
can be preserved by including the contributions from 
the Kaluza-Klein (KK) states of vector bosons \cite{unitary5D,Csaki:2003dt}.
It was found that by assuming 
the delocalized fermion profile \cite{Cacciapaglia:2004rb,Foadi:2004ps,
Cacciapaglia:2005pa,Chivukula:2005bn,Casalbuoni:2005rs}
or ideally delocalized fermion profile \cite{SekharChivukula:2005xm},
Higgsless models can be consistent with precision data. 
With this assumption, KK resonances in the model are fermiophobic and can escape all
experimental bounds on $W'$ and $Z'$. To discover the signature of the Higgsless models
with ideally delocalized fermions, vector boson scattering processes will be the key
detection means. As pointed out by Ref. \cite{Birkedal:2004au,He:2007ge}, the smoking gun for the
Higgsless model at LHC is a narrow resonance of ${\bar W}$ (${\bar W}$ is the next KK resonance of 
the charged vector boson, which can understood as the $W^R$ in the left-right hand model) in the $W Z$ channel.
Therefore, it is useful 
to examine spectra and the couplings of KK resonances in the Higgsless model 
to the vector bosons of the SM.
In this paper, we will examine both charged and neutral vector excitations as well as
the triple gauge couplings of ${\bar Z} WW$ (${\bar Z}$ is the
the next KK resonance of the neutral vector boson, which can be understood as 
the $Z^R$ in the left-right hand model) and $Z{\bar W} W$ vertices, 
which might be useful for phenomenological study on the searching for new
spin-1 resonances \cite{Davoudiasl:2003me,Birkedal:2004au,
Malhotra:2006sx,Piai:2007ys,He:2007ge}.

The multi-gauge boson vertices of Higgsless models with bulk gauge symmetry
$SU(2)\times SU(2)$ have been studied in both the
flat and warped space-time metrics, as shown in Ref. \cite{Chivukula:2005ji}. 
Triple gauge boson vertex $ZWW$ is also studied in
the Gauge-Higgs unification models, as shown in \cite{Sakamura:2006rf}. Although
the analytic expressions are provided in these works, the study on the
strength of KK vector bosons couplings to the 
SM vector bosons still lacks. This work is to provide a comprehensive
study on the spectra of new particles and the couplings of the neutral and charged KK vector 
bosons to the vector bosons of the SM. 

This work is also motivated by the work of Hirn and Sanz \cite{Hirn:2006nt},
where the S parameter can be negative in some region of parameter space
and the vector and axial gauge bosons may feel different effective metrics
by including a higher dimensional left-right kinetic mixing term.
Therefore it is interesting
to study the effects of these effective metrics
to mass spectra as well as triple and
quartic gauge couplings.
Recently there are also some controversies about the sign
of the S parameter in Hirn and Sanz model.
It is claimed that S is positive in this model
if the contribution of higher-dimensional operators
is subleading~\cite{Agashe:2007mc}.
This does not contradict with Hirn and Sanz model,
because the left-right kinetic mixing term is important
to induce different effective metrics for vector and axial sector.

Therefore, in this paper by assuming ideally delocalized fermion 
sector \cite{SekharChivukula:2005xm} and 
by including a gauge kinetic mixing term to the Higgsless model
(which we call Hirn-Sanz scenario),
we derive the equation of motion (EOM), boundary conditions (BC)
and formalism of multi-gauge-boson vertices. 
By using the shooting method, we study mass spectra as well as
the triple and quartic gauge couplings of vector bosons. 
We find that the deviation from the prediction of the SM in the triple gauge
couplings can reach $\pm 10\%$ by parameter $r_r$ and $\pm 5 \%$ by parameter 
$O_X$, both of which is within the reach of LHC and ILC.
We also find the triple gauge couplings of $Z{\bar W}W$ 
can be $ 50 \%$ smaller than the unitarity bounds.
While the triple gauge couplings of ${\bar Z}WW$ 
is $20 \%$ smaller than the unitarity bounds, which might
challenge the detection of ${\bar Z}$ via s channel at LHC.

The paper is organized as follows. In section II, 
the Higgsless model in Hirn-Sanz scenario is briefly reviewed. 
In section III, the formulas
for triple and quartic gauge couplings in Higgsless model are provided. 
In section IV,  numerical results are presented. We end this paper with
conclusions and discussion.

\section{Higgsless model in Hirn-Sanz Scenario}
Higgsless models are defined on the AdS$_5$ spacetime background. The metric of the
spacetime reads
\bea
ds^2 = \frac{R^2}{z^2} \sBBkf{ \eta_{\mu\nu} dx^{\mu} dx^{\nu} + dz^2 }\,,
\eea
where the $z$ is defined on the interval $[R, R']$. The UV brane
is located at $z=R$, and the IR boundary is located at $z=R'$. 
For fermion sector, we assume the ideally delocalized fermion \cite{SekharChivukula:2005xm},
then the fifth profile of the SM fermions 
can extend in the bulk and
guarantee the S parameter small enough and
the couplings to the KK resonances to vanish.
Since we consider the electroweak symmetry breaking,
below we drop the $SU(3)_c$ gauge 
group of the SM.
Weak gauge bosons are defined in the bulk, and can be interpreted as
the composite objects.

The Lagrangian with $SU(2)_L\times SU(2)_R\times U(1)_{B-L}$
gauge group in 5D reads as
\bea
{\cal  L} &=& \int d^4 x dz  \sqrt{g}  \sBBkf{
-\frac{1}{2 g_L^2(z)} Tr \mbkf{W_{MN,L} W^{MN}_L} -\frac{1}{ g_{LR}^2(z)} Tr \mbkf{W_{MN,L} \Omega^\dagger W^{MN}_R \Omega}  \nnb \\
&&-\frac{1}{2 g_R^2(z)} Tr \mbkf{W_{MN,R} W^{MN}_R} 
-\frac{1}{4 g_B^2(z)}    B_{MN} B^{MN}
}\,,
\label{lag}
\eea
where the indices $M, N=(\mu,5)$. 
The parameters $g_L$, $g_R$, and $g_B$ are gauge couplings of $SU(2)_L$, $SU(2)_R$
and $U(1)_{B-L}$, respectively. These couplings have mass dimension $-1/2$.
$W_{MN,L}$, $W_{MN,R}$, and $B_{MN}$ are strength tensors of 
the gauge fields. $\sqrt{g}$ is the Jacobian factor of the 5D space-time.
$\Omega$ is an auxiliary field, which transforms as a bifundamental field under
the gauge groups $SU(2)_L \times SU(2)_R$ and plays a role as the nonlinearly
realized 5D Goldstone boson~\cite{Hirn:2006nt}.
The gauge kinetic term can also be realized
in the linearly realized Higgs fields by a dimension-$6$ operator. 
However, in this article, we neglect the kinetic term of $\Omega$ field 
and the bulk mass terms of vector bosons.

In the Hirn-Sanz scenario, the gauge couplings can be parametrized as
\bea\frac{1}{g_{L}^2} =&\,\,\,\,\frac{ f_V + f_A } {2 g_0^2 }\,, \\
\frac{1}{g_{R}^2} =& r_{r}  \frac{ f_V + f_A } {2 g_0^2 }  \,,\\
\frac{1}{g_{LR}^2} =&  \frac{ f_V - f_A } {2 g_0^2 }  \,,\\
\frac{1}{g_{B}^2} =& r_{b}  \frac{ f_V + f_A } {2 g_0^2 } \,,
\label{eq:parameterization}
\eea
where $f_X$ are defined as
\bea
f_{X} &=& \exp \sBBkf{ \frac{O_X }{2 d (d-1)} z^{2d}  }\,,
\eea
where $X=V$ and $A$. For the sake of simplicity, we use universal parameters $O_X$
to describe gauge couplings $1/g_{L}^2$, $1/g_{LR}^2$, $1/g_{R}^2$, and $1/g_{B}^2$.
Parameters $O_V$ and $O_A$ are introduced to describe the QCD power corrections for the
two-point functions, and they can be related to the gluonic and fermionic condensates
\cite{Hirn:2006nt}. In the Higgsless model, parameters $O_V$ and $O_A$ can be interpreted
as the techni-gluonic and techni-fermionic condensates. 
The parameters $O_V$ and $O_A$ make it possible to study the 
cases where gauge couplings can have the fifth dimensional dependence,
and they can be either positive or negative.
According to the study of Ref. \cite{Karch:2006pv},
these terms modify the QCD OPE behavior.
The parameter $g_0$ is a dimensional constant to 
measure the strength of gauge couplings in 5D, 
which can be fixed by the normalization conditions of photon and W boson
as well as by fixing the coupling $g_{\gamma WW}=e$.

We introduce $r_r$ and $r_b$ to describe 
the relative strength of gauge couplings, respectively. 
Here we simply assume that $r_r$ and $r_b$ are independent of 
the fifth dimension.
The parameters $r_r$ and $r_b$ are assumed to be positive to avoid ghost-like behavior. 
Then, the gauge couplings $1/g_{L}^2$, $1/g_{R}^2$, and $1/g_{B}^2$ 
are always positive for any value of $O_X$. 
While the coupling of kinetic mixing term $1/g_{LR}^2$ 
can be either positive or negative, depending on the values of $O_V$ and $O_A$. 
The $r_r$ corresponds to the $1/\kappa$ as introduced in \cite{Chivukula:2005ji}.

To derive EOM and BC, we adopt the unitary gauge
by setting the fifth component of the gauge fields to zero.
In the construction of
Higgsless models, the symmetry of the model is further broken by BC.
The boundary conditions at UV brane ($z=R=R' \exp \left ( -b/2 \right ) $) are given as
\bea
\frac{1}{g_L^2} \partial_z W_{\mu,L}^a + \frac{1}{g_{LR}^2} \partial_z W_{\mu,R}^a  =0\,,\,\, W_{\mu, R}^{1,2} =0 \nnb \\
\frac{1}{g_B^2} \partial_z B_{\mu,L} + \frac{1}{g_{LR}^2} \partial W_{\mu,L}^3 + \frac{1}{g_R^2} \partial W_{\mu,R}^3 =0 \,,\,\,B_{\mu,L} - W_{\mu,R}^3 =0\,,
\label{uvbc}
\eea
which break the $SU(2)_R\times U(1)_{B-L}$ to $U(1)_Y$.
The boundary conditions at IR brane ($z=R'$) are given as
\bea
\left ( \frac{1}{g_L^2} + \frac{1}{g_{LR}^2} \right ) \partial_z W_{\mu,L}^a  + 
\left ( \frac{1}{g_R^2} + \frac{1}{g_{LR}^2} \right ) \partial_z W_{\mu,R}^a =0\,, \nnb \\
\,\,W_{\mu,L}^a - W_{\mu,R}^a  =0 \,,\,\,\partial_z B_\mu=0\,,
\label{irbc}
\eea
which break the $SU(2)_L \times SU(2)_R $ to the custodial symmetry $SU(2)_D$. We notice that
the kinetic mixing term of vector boson affect the boundary conditions, as indicating 
by the terms proportional to $1/g_{LR}^2$. 
After imposing all these symmetry breaking boundary conditions, 
the remnant symmetry both in the bulk and on the boundaries
is the $U(1)_{\textrm{EM}}$ gauge symmetry.

Following the compactification and dimensional reduction procedure,
the charged vector bosons in 5D can be decomposed with the
corresponding charged massive vector fields in 4D,
\bea
W_{\mu,L}^\pm(x,z) &=& \sum_{k=1}^{\infty}\psi^{L_{\pm}}_k(z) W^{k\pm}_\mu(x)\,, \label {crdecomp} \\
W_{\mu,R}^\pm(x,z) &=& \sum_{k=1}^{\infty}\psi^{R_{\pm}}_k(z) W^{k\pm}_\mu(x)\,, \label {cldecomp}
\eea
where $W^{k\pm}_\mu(x)$ are W and ${\bar W}$ bosons and their KK excitations.
The neutral vector fields in 5D can be decomposed with the 
corresponding neutral massless and massive
vector fields in 4D, 
\bea
B_\mu(x,z) &=& \psi_\gamma^B(z) \gamma_\mu(x) + \sum_{k=1}^{\infty}\psi^{B}_k(z) Z^k_\mu(x), \label {nbdecomp}\\
W_{\mu,L}^3(x,z) &=& \psi_\gamma^{L_3}(z) \gamma_\mu(x) + \sum_{k=1}^{\infty}\psi^{L_3}_k(z) Z^k_\mu(x), \label {nldecomp}\\
W_{\mu,R}^3(x,z) &=& \psi_\gamma^{R_3}(z) \gamma_\mu(x) + \sum_{k=1}^{\infty}\psi^{R_3}_k(z) Z^k_\mu(x), \label {nrdecomp}
\eea
where $\gamma_\mu(x)$, $Z^k_\mu(x)$ are 4D 
fields of photon, Z boson, ${\bar Z}$ and their KK excitations.
To describe the profile of charged vector bosons in the fifth dimension, 
wave functions have two branches: $\psi^{L_{\pm}}_k(z)$ and $\psi^{R_{\pm}}_k(z)$.
For each neutral particle, wave functions have three branches: $\psi^{B}_k(z)$,
$\psi^{L_3}_k(z)$ and $\psi^{R_3}_k(z)$.

The wave function $\psi_k(z)$ are determined by the on-shell equation of motions. 
For the mode functions of charged W boson and their KK excitations, 
the equation of motion read
\bea
\partial^2 \psi_k^{L_\pm}
+ \sBBkf{ \partial \ln J_5 + c  \partial \ln \frac{1}{g_{L}^2} + (1 -c) \partial \ln\frac{1}{g_{LR}^2}  } \partial \psi_k^{L_\pm}
+ c \frac{g_{L}^2}{g_{LR}^2} \partial \ln \frac{g_{R}^2}{g_{LR}^2} \partial \psi_k^{R_\pm}
&=& - \frac{J_4}{J_5} M^2_{W_k} \psi_k^{L_\pm}\,,\label{psilw}\\
\partial^2 \psi_k^{R_\pm}
+ \sBBkf{ \partial \ln J_5 + c \partial \ln \frac{1}{g_{R}^2} + (1 -c) \partial \ln\frac{1}{g_{LR}^2}   } \partial \psi_k^{R_\pm}
+ c \frac{g_{R}^2}{g_{LR}^2} \partial \ln \frac{g_{L}^2}{g_{LR}^2} \partial \psi_k^{L_\pm}
&=& - \frac{J_4}{J_5} M^2_{W_k} \psi_k^{R_\pm} \,.\label{psirw}
\eea
For the mode functions of neutral vector bosons, the equation of motion read as
\bea
\partial^2 \psi_k^{L_3}
+ \sBBkf{ \partial \ln J_5 + c \partial \ln \frac{1}{g_{L}^2} + (1-c)  \partial \ln\frac{1}{g_{LR}^2}  } \partial \psi_k^{L_3}
+ c \frac{g_{L}^2}{g_{LR}^2} \partial \ln \frac{g_{R}^2}{g_{LR}^2} \partial \psi_k^{R_3}
&=&- \frac{J_4}{J_5} M^2_{Z_k} \psi_k^{L_3}\,,\label{psil3}\\
\partial^2 \psi_k^{R_3}
+ \sBBkf{ \partial \ln J_5 + c \partial \ln \frac{1}{g_{R}^2} + (1-c)  \partial \ln\frac{1}{g_{LR}^2}  } \partial \psi_k^{R_3}
+ c \frac{g_{R}^2}{g_{LR}^2} \partial \ln \frac{g_{L}^2}{g_{LR}^2} \partial  \psi_k^{L_3}
&=&-\frac{J_4}{J_5} M^2_{Z_k} \psi_k^{R_3}\,,\label{psir3} \\
\partial^2 \psi_k^{B}
+ \sBBkf{ \partial \ln J_5 +   \partial \ln\frac{1}{g_{B}^2} } \partial \psi_k^{B}
&=&- \frac{J_4}{J_5} M^2_{Z_k} \psi_k^{B} \,, \label{psib3}
\eea
with
\bea
\frac{1}{c} & =& 1 - \frac{g_{L}^2 g_{R}^2}{g_{LR}^4}\,.
\eea
To simplify the expression, we introduce a shorthand convention 
\bea
\partial \ln \frac{1}{g_{i}^2} & \equiv & - \frac{ \partial g_{i}^2}{g_{i}^2} \,.
\eea
which is independent of the sign of $g_{i}^2$, with $(i=L, R, B, LR)$, which is also
true for $J_5$.

The $J_4$ and $J_5$ are defined as
\bea
J_4 &=& \sqrt{g} g^{\mu\mu'} g^{\nu\nu'}\,,\\
J_5 &=& \sqrt{g} g^{\mu\mu'} g^{55}\,.
\eea
For AdS$_5$ case, we have
\bea
J_4 & =&  \,\,\,\,  R' \frac{\exp\{-b/2\}}{z},\\
J_5 & =&  - R' \frac{\exp\{-b/2\}}{z}\,.
\eea

We observe that the operator $Tr\mbkf{W_{MN,L}\Omega^\dagger W^{MN}_R\Omega}$ induces
the entangling terms in the Eqs. (\ref{psilw}-\ref{psib3}), which are the last terms in the
left-hand side of these equations. When $1/g_{LR}^2$ (or $f_V = f_A$) equal zero, 
the contributions of kinetic mixing terms to boundary conditions as well as EOMs vanish.

Another fact is that the constant $g_0$ does not affect
either EOM or boundary conditions, which means that it does not affect the 
mass spectra of vector bosons. 

By using the boundary conditions of symmetry breaking given in 
Eqs. (\ref{uvbc}-\ref{irbc}) and Eqs. (\ref{cldecomp}-\ref{nrdecomp}),
the UV boundary conditions for wave function $\psi$ at $z=R$ are 
determined by Eq. (\ref{uvbc}) as
\bea
\left ( \frac{1}{g_L^2} \partial_z \psi^{L_\pm}_n + \frac{1}{g_{LR}^2} \partial_z \psi^{R_\pm}_n \right ) |_{z=R} = 0,
\psi^{R_\pm}_n |_{z=R} =0, \nnb \\
( \frac{1}{g_R^2} \partial_z \psi^{R_3}_n  + \frac{1}{g_{LR}^2} \partial_z \psi^{L_3}_n + \frac{1}{g_B^2} \partial_z \psi^B_n) |_{z=R} =0,
( \psi^{R_3}_n  - \psi^B_n ) |_{z=R}=0\,.
\label{bcuv}
\eea
And the IR boundary conditions for wave function $\psi$ at $z=R'$  are determined by
Eq. (\ref{irbc}) as
\bea
\left [ (\frac{1}{g_L^2} + \frac{1}{g_{LR}^2}) \partial_z \psi^{L_\pm,L_3}_n  
+ ( \frac{1}{g_R^2} + \frac{1}{g_{LR}^2}) \partial \psi^{R_\pm,R_3}_n \right ] |_{z=R'} = 0, \nnb \\
(\psi^{L_\pm,L_3}_n - \psi^{R_\pm,R_3}_n)|_{z=R'} =0,
\partial_z \psi^B_n |_{z=R'}=0\, .
\label{bcir}
\eea

We observe that 
if the gauge couplings have non-trivial dependence on $z$,
they can affect both boundary conditions and equation of motions of KK modes
in a non-trivial way. 

Since there is an extra $U(1)_{B-L}$ sector in the Higgsless model and there is no 
extra advantage for formulating two-, three-, and four-point functions by using the
vector and axial vector fields, 
we use the left and right vector fields in both EOM and boundary conditions. 
However, we find that Eqs. (\ref{psilw}-\ref{psib3}) as well as Eqs. (\ref{bcuv}-\ref{bcir}) 
can reduce to those in vector and axial-vector fields when the gauge symmetry is 
taken as $SU(2)_L \times SU(2)_R$ and $g_{L}=g_R$ is imposed.
In the original Hirn-Sanz scenario, the QCD corrections can be interpreted as that 
vector and axial gauge bosons can feel different metrics. Here we adopt the 
interpretation of the QCD corrections as the modification of the gauge 
couplings $1/g_{L}^2$, $1/g_{LR}^2$, and $1/g_{R}^2$, as
shown in the EOM and boundary conditions given above. 

The normalization conditions for wave functions of neutral 
and charged bosons are given as
\bea
\int_z J_4 \sBkf{\frac{ \psi^{L_\pm}_n \psi^{L_\mp}_n } {g_L^2(z)} 
+ 2 \frac{ \psi^{L_\pm}_n \psi^{R_\mp}_n } {g_{LR}^2(z)} 
+ \frac{ \psi^{R_\pm}_n \psi^{R_\mp}_n } {g_R^2(z)} } &=& 1\,, \\
\int_z J_4 \sBkf{\frac{ \psi^{L_3}_n \psi^{L_3}_n } {g_L^2(z)} 
+ 2 \frac{ \psi^{L_3}_n \psi^{R_3}_n } {g_{LR}^2(z)} 
+ \frac{ \psi^{R_3}_n \psi^{R_3}_n } {g_R^2(z)}
+ \frac{\psi^B_n \psi^B_n }{g_B^2(z)} } &=& 1\,, \label{norm-z}
\label{norm-w}
\eea
which guarantee the canonically normalized kinetic terms of vector bosons.
In our numerical method, we choose to find eigen-wavefunctions with the shooting method 
from IR brane to UV brane. We need to choose an initial value of $\psi^L_n$ 
as input in order to determine the whole wavefunctions. The normalization 
condition is essential to remove the arbitrariness in the magnitude of $\psi^L_n$. 
However, there still exists an ambiguity on the sign of $\psi^L_n$. Therefore 
we always choose a positive $\psi^L_n$ in our numerical calculation.

\section{Formalism for multi-gauge boson vertices}

The triple gauge couplings are induced from the non-Abelian structure of 
the model in 5D. Below we consider the
couplings of W pair to neutral vector bosons,
which can be expressed straightforwardly into two terms:
\bea
- i k_{NWW} N_{\mu\nu} W_+^{\mu} W_-^{\nu} 
- i g_{NWW} (N_{\mu} W_{\nu, -} W_+^{\mu\nu} - N_{\mu} W_{\nu, +} W_-^{\mu\nu})\,.
\eea
These two terms are part of the more general 
parametrization of triple gauge couplings \cite{Hagiwara:1986vm}, since we have not included
higher dimensional operators nor the operators violating $C$ and $CP$ symmetries
in the original Lagrangian given in Eq. (\ref{lag}).

The effective couplings $k_{NWW}$ and $g_{NWW}$ are determined
by the integral over the fifth dimension as
\bea
\int_z J_4 \sBkf{\frac{ \psi^{L_\pm}_W \psi^{L_\mp}_W \psi^{L}_N} {g_L^2(z)} 
+ \frac{ \psi^{L_\pm}_W \psi^{L_\mp}_W \psi^{R}_N} {g_{LR}^2(z)}
+ \frac{ \psi^{R_\pm}_W \psi^{R_\mp}_W \psi^{L}_N} {g_{LR}^2(z)}
+ \frac{ \psi^{R_\pm}_W \psi^{R_\mp}_W  \psi^{R}_N } {g_R^2(z)} } &=& k_{NWW}\,,\\
\int_z J_4 \sBkf{\frac{ \psi^{L_\pm}_W \psi^{L_\mp}_W \psi^{L}_N} {g_L^2(z)} 
+ \frac{ \psi^{L_\pm}_W  \psi^{L}_N \psi^{R_\mp}_W} {g_{LR}^2(z)}
+ \frac{ \psi^{R_\pm}_W  \psi^{R}_N \psi^{L_\mp}_W} {g_{LR}^2(z)}
+ \frac{ \psi^{R_\pm}_W \psi^{R_\mp}_W  \psi^{R}_N } {g_R^2(z)} } &=& g_{NWW}\,.
\eea
In order to fix the free parameter $g_0$, we introduce the charge universality condition 
\bea
g_{\gamma WW} \equiv e \,.
\label{charge-uni}
\eea
We find vertices of photon with charged vector bosons are the same, 
which is expected from the symmetry breaking pattern, 
as demonstrated by the first column in Table IV and Table V.
We can extend the above formalism straightforward to 
study the triple gauge couplings of ${\bar W}$ to all neutral vector bosons.

The quartic gauge couplings are also induced from the non-Abelian structure of the Higgsless
model in 5D. The quartic gauge couplings of $WWWW$ can be cast into two terms
\bea
- g_{WWWW}^2 \left ( W^+ \cdot W^+  W^- \cdot W^-  -  W^+ \cdot W^-  W^+ \cdot W^- \right )\,,
\eea
where the coupling of $WWWW$ is defined as
\bea
\int_z J_4 \sBkf{
   \frac{ |\psi^{L_\pm}_W \psi^{L_\pm}_W|^2} {g_L^2(z)} 
+2 \frac{ |\psi^{L_\pm}_W \psi^{R_\mp}_W|^2} {g_{LR}^2(z)}
+  \frac{ |\psi^{R_\pm}_W \psi^{R_\mp}_W|^2} {g_R^2(z)} } &=& g_{WWWW}^2\,.
\eea
It is easy to extend the coupling $g_{WWWW}^2$ to $g_{{\bar W} {\bar W} {\bar W} {\bar W}}^2$.

The quartic gauge couplings of $ZZWW$ can be cast into two terms
\bea
- g_{ZZWW}^2 \left ( Z \cdot Z  W^+ \cdot W^-  -  Z \cdot W^+  Z \cdot W^- \right )\,,
\eea
where the coupling of $ZZWW$ is defined as
\bea
\int_z J_4 \sBkf{
   \frac{ (\psi^{L_3}_Z)^2 \psi^{L_\pm}_W \psi^{L_\mp}_W} {g_L^2(z)} 
+2 \frac{ \psi^{L_3}_Z \psi^{R_3}_Z \psi^{L_\mp}_W \psi^{R_\pm}_W} {g_{LR}^2(z)}
+  \frac{ (\psi^{R_3}_Z)^2 \psi^{R_\pm}_W \psi^{R_\mp}_W } {g_R^2(z)} } &=& g_{ZZWW}^2\,.
\eea
We neglect other quartic gauge couplings, like ${\bar Z} {\bar Z} WW$ in this paper.

We observe that the $U(1)_{B-L}$ gauge kinetic term 
does not contribute three-point nor four-point vertices, due to its Abelian nature and
there is no self-interaction terms in its gauge kinetic term.

In warped spacetime, since all vertices are determined by the parameters of the model, 
we can examine the unitarity violation
for the $WW$ scattering process by introducing the following two parameters:
\bea
\delta U_{E^4}^W \equiv   g_{WWWW}^2 -     \sum_N g_{NWW} g_{NWW}\,, \\
\delta U_{E^2}^W \equiv 4 g_{WWWW}^2 - 3   \sum_N g_{NWW} g_{NWW} \frac{ M_N^2}{M_W^2} \,.
\eea
$U_{E^4}$ determines the magnitude of the term of $E^4/m_W^4$,
and $U_{E^2}$ determines the magnitude of the term of $E^2/m_W^2$. 

\section{Numerical Analysis}
Before resorting to numerical solution, it is instructive to count
the number of free parameters in the model.
We have $6$ free parameters in total: $b$, $r_{r}$, 
$r_{b}$, $O_V$, $O_A$, and $d$. In the following study, 
we fix $b=24$ and $d=2$ for simplicity 
of the numerical analysis. $b=24$ corresponds to 
the UV cutoff  set at $10^4$ TeV or so.

We use the following experimental data to determine eigenvalues and 
eigen-wavefunctions as well as the TGC of $g_{\gamma W W}$:
\bea
m_Z = 91.19 \textrm{GeV}, \,\, e = 0.3124, \,\, \sin^2\theta_W = 0.2315\,.
\eea

We can organize the spectra of vector bosons into a quintuplet and its excitations:
a charged doublet vector bosons, $W^\pm$ and ${\bar W}^\pm$, and 
a triplet neutral vector boson, $\gamma$, $Z$, and ${\bar Z}$, which correspond to the
$U(1)^3$ subgroup of $SU(2)_L \times SU(2)_R \times U(1)$. We show the first quintuplet,
second, and third excitations by taking five points as examples:
For point 1, the kinetic
term does not contribute to the EOM or to the boundary conditions. The gauge 
couplings are set to be constant in the fifth dimension. We use 
this dependence of gauge couplings on fifth dimension as trivial case. 
This point will serve as 
the reference point to compare and contrast other points.

For point 2, the gauge
couplings have 5D profile while the gauge kinetic mixing term vanishes at tree-level. 
For  point 3, the gauge couplings have 5D profile and the
kinetic mixing term contributes to both the EOM and the 
boundary conditions. 
Points 4 and 5 are similar to  points 2 and 3.
The difference is that the gauge couplings increase with the increase of $z$ for 
 points 2 and 3, 
while the gauge couplings decrease with the increase of $z$ for  points 4 and 5.

To select these points, we impose $O_A > O_V$
to satisfy Witten's positivity condition \cite{Witten:1983ut}.
We deliberately choose $r_b$ to guarantee the
mass ratio of $M_W^2$ close to its experimental value. 
The large $r_b$  corresponds the small $g_{B}^2$.
 For these five point case study,
we fix $r_r=1$.
 
The parameter $\Lambda_{EW}$ is about $300$ GeV.
To determine the mass spectra, we fix the IR brane on
the point $z=1$ by choosing $\Lambda_{EW} R' = 1$ and the UV brane on the 
point $z=\epsilon_+=\exp\{-b/2\}$. By fixing the mass of z boson we can 
find the parameter $\Lambda_{EW}$. The value of 
$\sqrt{\Lambda_{\textrm{EW}}} g_0$ demonstrates these points indeed locates
at weakly coupled region. 

\begin{table}[ht]
\begin{center}
\begin{tabular}{| c|| c|| c| c|| c| c|}\hline
& & & & & \\
&$P_1$ & $P_2$  & $P_3$   & $P_4$ &  $P_5$   \\ 
\hline \hline
& & & & & \\
$r_b$ &$2.85$ & $2.850$ & $2.50$ & $2.45$ & $2.45$   \\
& & & & & \\ \hline
& & & & & \\
$O_V$ &$0.00$ & $10.00$ & $0.00$  & $-10.00$ & $-10.00$ \\
& & & &  & \\ \hline
& & & &  & \\
$O_A$ &$0.00$ & $10.00$ & $10.00$ & $-10.00$ & $0.00$ \\
& & & &  & \\ \hline \hline
& & & &  & \\
$\Lambda_{\textrm{EM}} (\textrm{GeV})$ &$275.24$ & $208.03$ & $201.72$ & $481.14$ & $271.81$ \\
& & & &  & \\ \hline
& & & &  & \\
$\sqrt{\Lambda_{\textrm{EM}}} g_0 \times 10^3 $ &$5.85$ & $6.24$ & $5.78$ & $5.57$ & $5.62$ \\
& & & &  & \\ \hline
\end{tabular}
\end{center}
\caption{ Selected points in parameter space to show spectra and couplings}
\label{tablebp}
\end{table}

\begin{table}[ht]
\begin{center}
\begin{tabular}{| c| c| c| c| c| c| c|}\hline
& & & & & & \\
&$W$ & $\overline{W}$  & $W^{\prime}$   & $\overline{W}^{\prime}$  & $W^{\prime\prime}$   & $\overline{W}^{\prime\prime}$\\ 
\hline
& & & & & & \\
$P_1$ &$80.68$ & $680.48$  & $1075.27$ & $1539.78$ & $1952.37$ & $2403.28$ \\
& & & & & & \\ \hline \hline
& & & & & & \\
$P_2$ &$80.22$ & $325.05$  & $931.00$ & $1096.08$ & $1563.76$ & $1776.53$ \\
& & & & & & \\ \hline
& & & & & & \\
$P_3$ &$80.36$ & $499.21$  & $903.47$ & $1129.11$ & $1516.71$ & $1761.84$ \\
& & & & & & \\ \hline \hline
& & & & & & \\
$P_4$ &$80.20$ & $1569.89$ & $1795.07$ & $3038.59$  & $3401.35$ & $4459.40$ \\
& & & & & & \\ \hline
& & & & & & \\
$P_5$ &$80.55$ & $888.50$ & $1063.81$ & $1716.73$  & $1929.39$ & $2519.27$ \\
& & & & & & \\ \hline
\end{tabular}
\end{center}
\caption{ The spectrum of charged vector bosons. The unit of mass is GeV.}
\label{tablewmass}
\end{table}

\begin{table}[ht]
\begin{center}
\begin{tabular}{|c| c| c| c| c| c| c| c| c| c|}\hline
& & & & & & & & & \\
&$\gamma$ & $Z$ &  $\overline{Z}$  &  $\gamma^{\prime}$  & $Z^{\prime}$ & $\overline{Z}^{\prime}$ & $\gamma^{\prime\prime}$ & $Z^{\prime \prime}$ & $\overline{Z}^{\prime\prime}$\\ 
\hline
& & & & & & & & &\\
$P_1$  &$0$ & $91.19$  & $675.48$ & $699.67$ & $1080.69$ & $1534.32$ & $1560.53$ & $1958.02$ & $2397.57$\\
& & & & & & & & & \\ \hline \hline
& & & & & & & & & \\ 
$P_2$  &$0$ & $91.19$  & $320.60$  & $340.57$ & $934.90$ & $1091.76$ & $1111.45$ & $1567.81$ & $1772.18$ \\ 
& & & & & & & & &\\ \hline
& & & & & & & & &\\ 
$P_3$ &$0$ & $91.19$  & $370.33$ & $503.56$ & $907.34$ & $1064.98$ & $1134.48$ & $1520.87$ & $1723.24$ \\ 
& & & & & & & & &\\ \hline \hline
& & & & & & & & &\\ 
$P_4$ &$0$ & $91.19$  & $1560.07$ & $1614.09$ & $1805.69$ & $3029.12$ & $3075.59$ & $3411.93$ & $4449.22$ \\ 
& & & & & & & & &\\ \hline
& & & & & & & & &\\ 
$P_5$ &$0$ & $91.19$  & $755.63$ & $896.75$ & $1068.71$ & $1557.74$ & $1723.85$ & $1934.74$ & $2397.04$ \\ 
& & & & & & & & &\\ \hline
\end{tabular}
\end{center}
\caption{ The spectrum of neutral vector bosons. The unit of mass is GeV. }
\label{tablezmass}
\end{table}

The mass spectra of vector
bosons are listed in Table II and Table III, for the neutral and charged 
vector bosons, respectively. Compared with the point 1, 
the point 2-like gauge coupling 
dependence can decrease the mass difference of vector bosons, while the point 4-like
gauge coupling dependence can increase the mass difference of vector bosons. 
Obviously, QCD-like corrections can modify the spectra dramatically.
The spectra of point 3 is similar to point 2, especially for
higher KK excitations. This means that the sign
of $1/g_{LR}^2$ can make differences for lower KK excitations.
The spectra of point 5 is quite different from those of point 4.
This means that the sign
of $1/g_{LR}^2$ can make differences for all KK excitations.

Table IIV lists the TGC of 
$W W N$, while Table IV lists the TGC of ${\bar W} {\bar W} N$.
In point 1, we can have the exact identity $g_{NWW}=k_{NWW}$. 
In other points, such an identity is broken negligibly,
which can be understood simply from Eq.(31-32),
where only $g_{LR}$ terms have some minor difference
and left and right gauge boson profiles difference is not big
enough to break the identity heavily.
Therefore we only list $g_{NWW}$ type TGC.

The coupling $g_Z$ in the SM is given as
$0.5786$, the deviation can reach to
$+4.4\%$, $4.3\%$, $+2.7\%$, $+0.3 \% $, and $+1.3 \% $ 
for each points, respectively.
Such deviations can saturate the LEP2 TGC measurement bounds.
We observe that $W$ pair has larger couplings with photon and Z boson
than other neutral vector bosons. 
Couplings of $NWW$ to higher KK excitation decrease quickly with 
the increase of KK number. 

The pattern of the couplings of ${\bar W}$ pair to neutral vector bosons 
becomes more complicated. There are several comments in order.
1) The charge universality is guaranteed, which is realized when only we include
kinetic term consistently in both equations of motion and boundary conditions.
This serves as one crucial check for our numerical approach.
2) Generally speaking, the ${\bar W}$ pair strongly couple to ${\bar Z}$ and $\gamma'$.
3) The tendency of that the couplings 
of the ${\bar W}$ pair to neutral bosons decrease when KK number of neutral vector 
bosons goes to higher is also demonstrated
by Table V, though not as quickly as the cases for $W$ pair.

Fig. 1 is to show the dependence of $ZWW$ on the parameter $O_X$ and $r_r$.
We select three cases with 1) $O_V=O_A=O_X$, 
2) $O_V=O_X, O_A=0$, and 3) $O_V=0, O_A=O_X$ to show the dependence of
$g_{ZWW}$ on $O_X$. In Fig. 1, $r_r$ is fixed to $1$.
We follow \cite{Hirn:2006nt} to vary $O_X$ from $-15$ to $15$.
There are several comments in order for Fig. (1a): 
(1) When $O_X$ is smaller than $0$, the
difference in $g_{ZWW}$ between these three cases is small. While when
$O_X$ is larger than $0$, with the increase of $O_X$, 
$g_{ZWW}$ decreases in case 1) and case 2) while is increase in case 3).
The $O_X$ can modify the $g_{ZWW}$ up to $\pm 5\%$.
(2) We fix $O_A=O_V$ and select three cases 
to show the effects of $O_X$ while $r_r$ can vary from $0.07$ to $8.89$:
1) with $O_X=0$, 2) with $O_X=10$, and 3) with $O_X=-10$.
It is obviously that the $g_{ZWW}$ can be either larger or smaller 
than the prediction of the SM. 
The deviation from the prediction of the SM can even 
reach to $100 \%$ when $r_r$ is larger than $8$. 
To obtain this large deviation we fix $r_b=1$, which cannot guarantee that 
the $m_W$ equals to its experimental value. But after adjusting $r_b$ to make
$m_W$ equal to its experimental value, we find the deviation can be larger than 
a few $10 \%$.

The unitary bounds of the coupling ${\bar Z}WW$
can be put as \cite{Malhotra:2006sx}
\bea
u_{{\bar Z} W W} & = & g_{Z W W} \frac{m_Z}{\sqrt{3} M_{\bar Z} } \,,
\eea
The unitary bounds of the coupling $g_{Z W {\bar W}}$ 
can be put as \cite{Birkedal:2004au}
\bea
u_{Z {\bar W} W } &=& g_{Z W W} \frac{m_Z^2}{\sqrt{3} M_{\bar W} M_W } \,,
\eea
To show the deviation from this unitarity estimate, we define
\bea
r_{{\bar Z} W W} &=& \frac{ g_{{\bar Z} W W}} {u_{ {\bar Z} W W}}\,, \\
r_{Z {\bar W} W} &=& \frac{ g_{Z {\bar W} W}} {u_{Z {\bar W} W}}\,.
\eea

Fig. 2 shows the dependence of $r_{{\bar Z} WW}$ and $r_{Z{\bar W} W}$ 
on $O_X$ and $r_r$. In Fig. 2 we select three cases with 1) $O_V=O_A=0$, 
2) $O_V=O_A=10$, and 3) $O_V=O_A=-10$ to show the dependence of
$g_{ZWW}$ and $g_{Z{\bar W} W}$ on $r_r$.
Fig. 2a) shows that  $r_{{\bar Z} WW}$ is smaller than $0.25$ for case 2),
$12 \%$ for case 1), and $0.035 \%$ for case 3). The underlying reason for 
smallness of $r_{{\bar Z} WW}$ is related with the
almost degeneracy of ${\bar Z}$ and $\gamma'$, as demonstrated in Table III and
Table IV.
Fig. 2b) shows that  $r_{Z {\bar W}W}$ is closer to its unitarity bound value
than $r_{{\bar Z} WW}$ does. When $r_r$ is close $1$, the value of $r_{{\bar Z} WW}$ 
and $r_{Z{\bar W} W}$ approach their maximum values.

\begin{figure}[t]
\centerline{
\includegraphics[width=8cm]{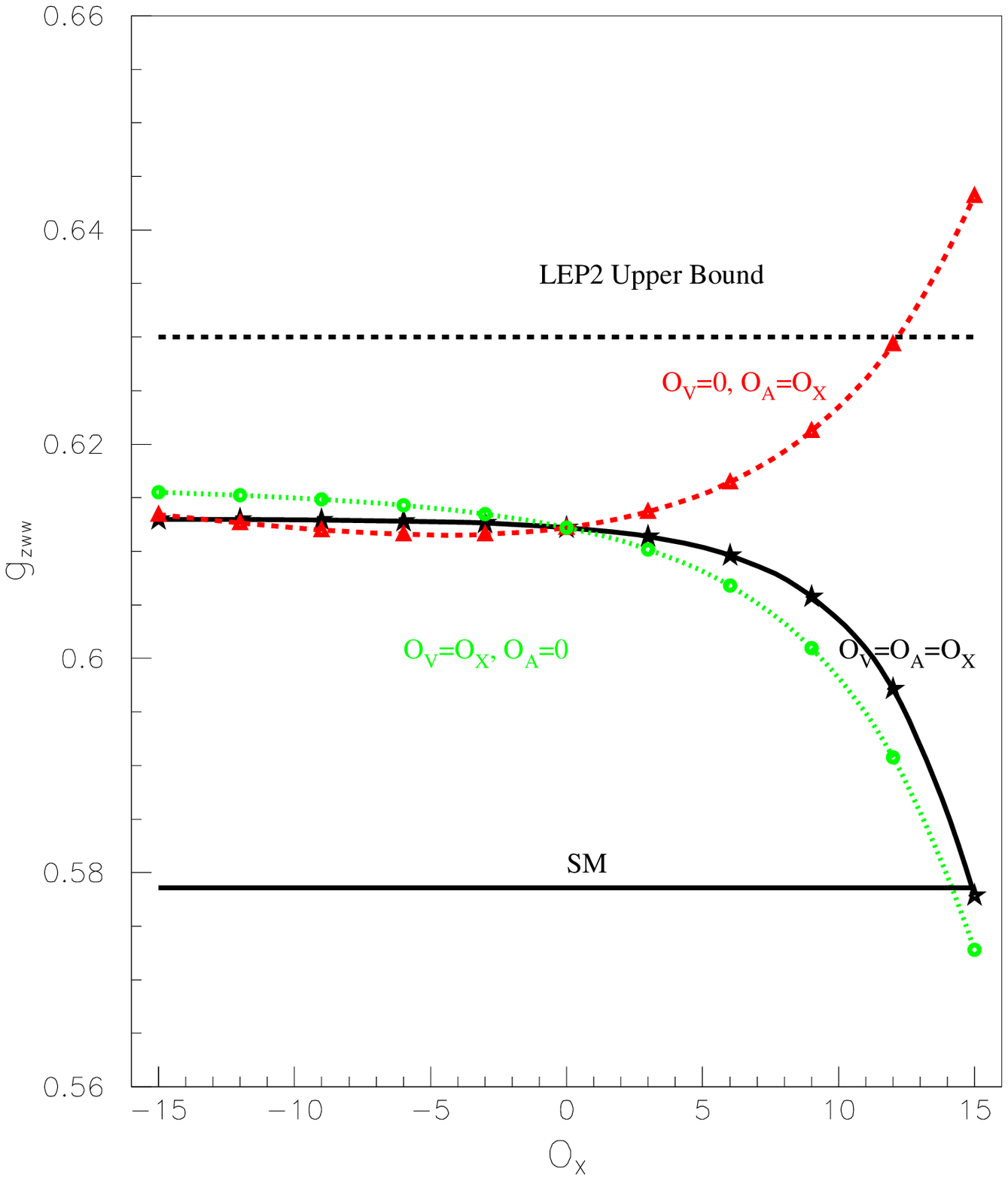}
\hspace*{0.0cm}
\includegraphics[width=8cm]{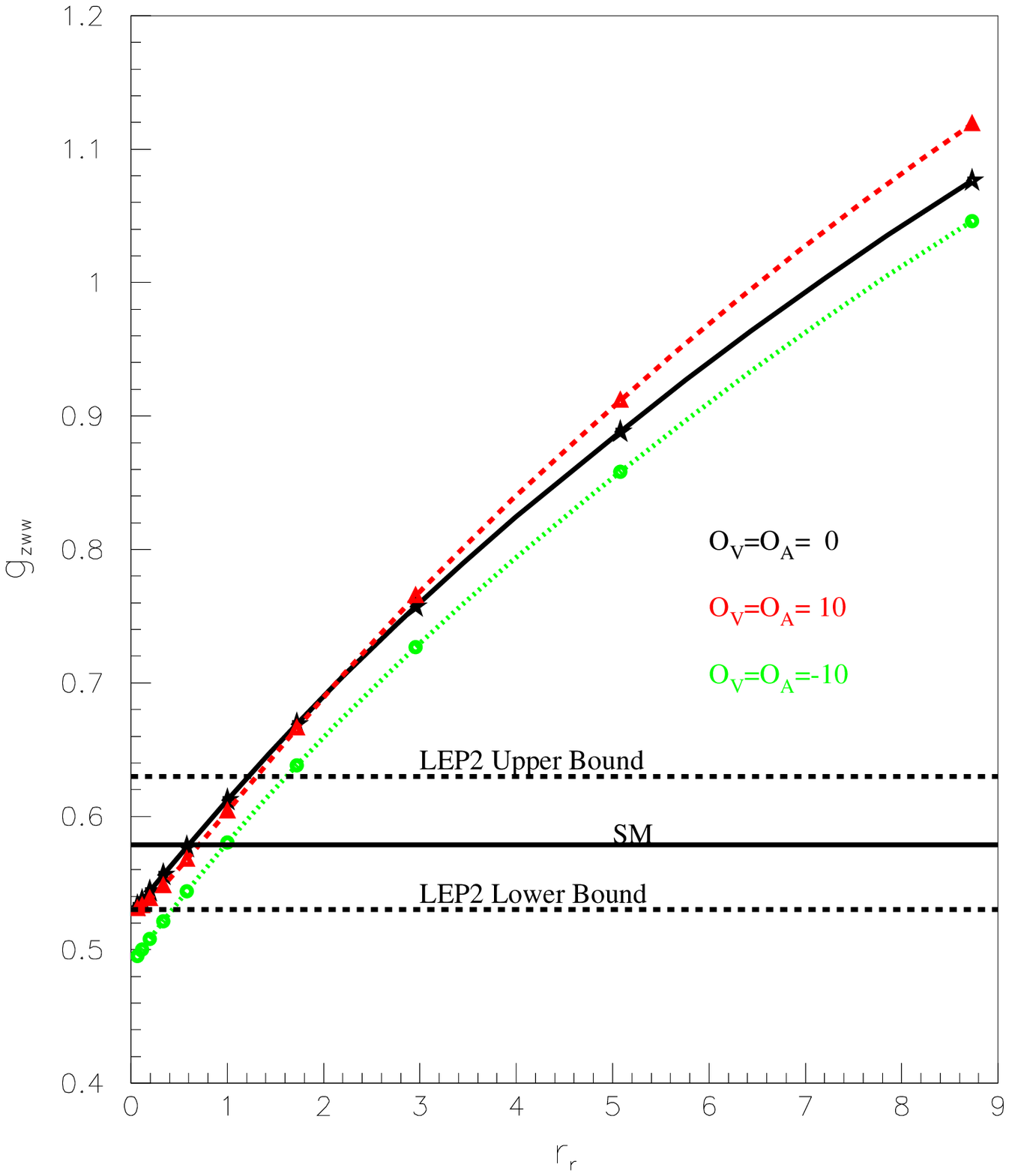}
}
\vskip -.25cm
\hskip 0.8 cm {\bf ( a) } \hskip 7.5cm {\bf (b)}
\caption{\it
Triple gauge couplings, $g_{ZWW}$, vary with $O_X$ and $r_r$.
In fig. 1a), we select three cases to show the effects of $O_X$: 1) $O_V=O_A=O_X$,
2) $O_V=O_X, O_A=0$, and 3) $O_V=0, O_A=O_X$.
In fig. 1b), we select three cases to show the effects of $r_r$: 1) $O_V=O_A=0$, 
2) $O_V=O_A=10$, and 3) $O_V=O_A=-10$. For both fig. 1a) and 1b), $r_b$ is fixed as $1$.
In figure 1a, the LEP2 two parameter fit upper bound is plotted, while in 
figure 1b, both the upper and lower bounds are plotted. Future colliders' bounds on $g_{ZWW}$
can reach $10^{-3}$ and are not depicted.}
\label{fig1}
\end{figure}

The deviation of QGC for four $W$ vertex is about $+6.7\%$, $+8.0\%$, 
$+4.2\%$,$-0.3\%$, and $+1.5\%$,  from the value of SM, $0.4356$.
The deviation of QGC for $ZZWW$ vertex is about $+9.9 \%$, $+12.5 \%$, 
$+6.7\%$, $+0.6\%$, and $+3.0\%$, from the value of SM, $0.3348$. 
It is remarkable that the QGC of four ${\bar W}$ are 
larger than $1.2$, as shown in Table V.

Fig. 3 shows the quartic gauge couplings $g_{WWWW}^2$ 
and $g_{{\bar W}{\bar W}{\bar W}{\bar W}}^2$ versus $O_X$.
It is obvious that the quartic gauge coupling $g_{WWWW}^2$ is almost unchanged
when $O_X$ is smaller than $0$, while it changes when $O_X$ is larger than $0$.
Such a behavior is similar to the case of $g_{ZWW}$, which can be interpreted
as the consequence of unitarity.
The $g_{{\bar W}{\bar W}{\bar W}{\bar W}}^2$ decreases steeply with the increase of
$O_X$ for case 1) and case 2); while increase mildly with the increase of $O_X$ for case 3).
The coupling $g_{{\bar W}{\bar W}{\bar W}{\bar W}}^2$ can be larger than one due to the fact
that the wavefunctions of ${\bar W}$ can be larger than one. 
In order to guarantee the interactions of four ${\bar W}$ is weak, Fig. 3b implies that
$O_X$ should be larger than $10$ or so.

\begin{figure}[t]
\centerline{
\includegraphics[width=8cm]{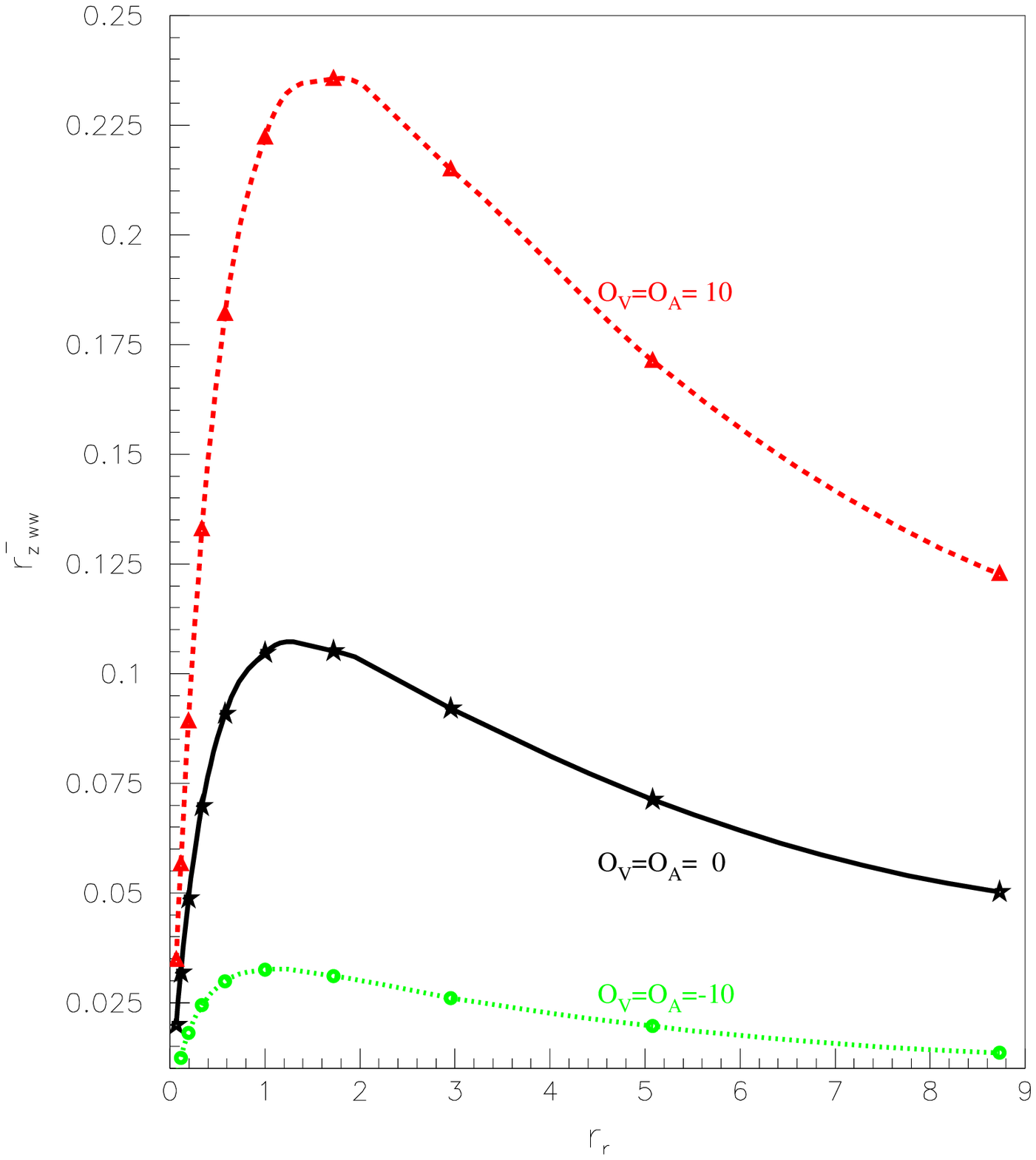}
\hspace*{0.0cm}
\includegraphics[width=8cm]{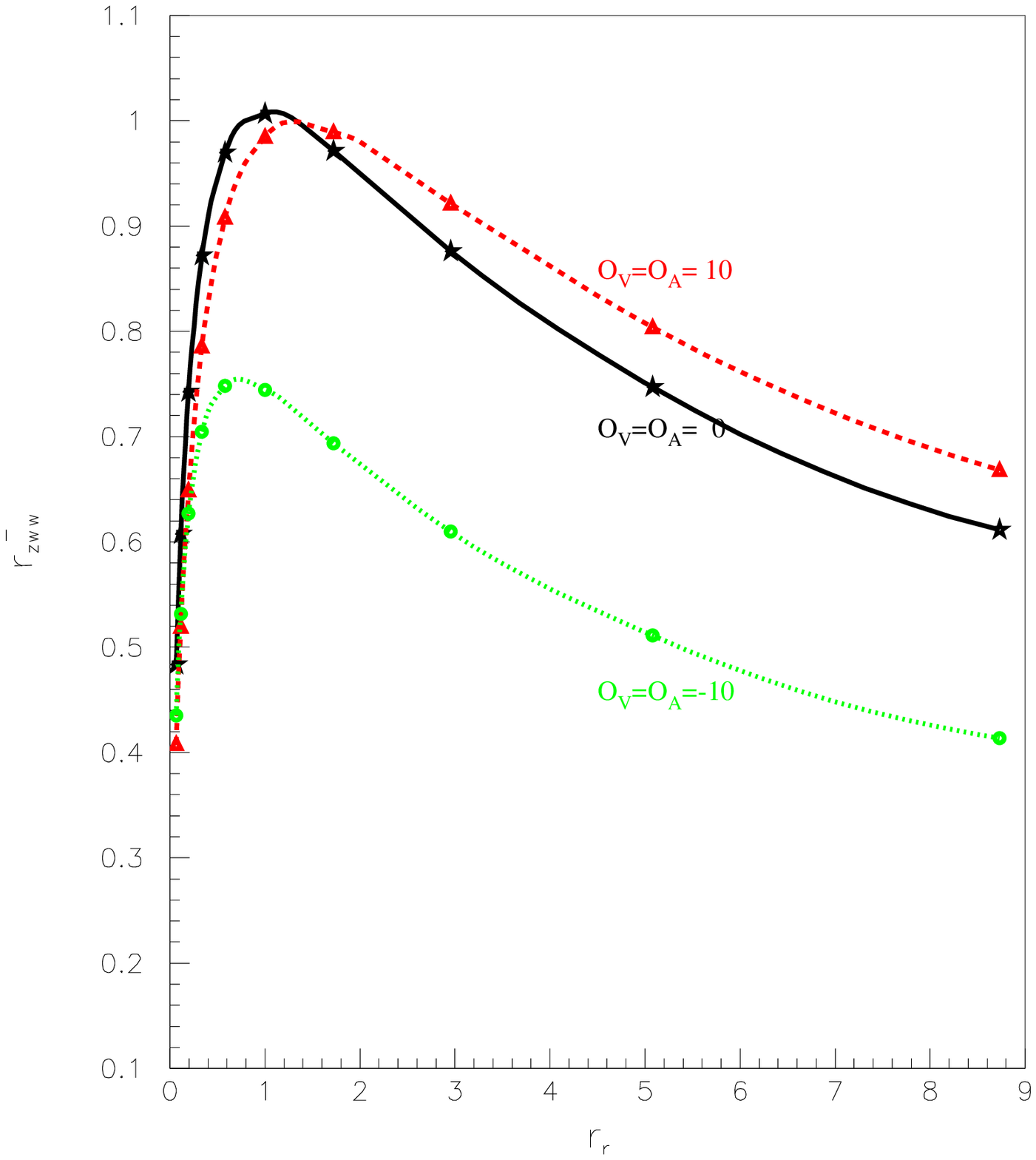}
}
\vskip -.25cm
\hskip 0.8 cm {\bf ( a) } \hskip 7.5cm {\bf (b)}
\caption{\it
The ratio of triple gauge couplings, $r_{{\bar Z} WW}$ and $r_{Z{\bar W} W}$, 
vary with $r_r$ from $0.07$ to $8.89$. We fix $O_V=O_A$ and select three cases to 
show the effects of $O_X$: 1) $O_X=0$, 2) $O_X=10$, and 3) $O_X=-10$. Here the $r_b$ is fixed
to $1$.
}
\label{fig2}
\end{figure}

It is impossible to sum infinite number of KK resonances 
to examine the unitarity violation. In the
realistic calculation, we sum neutral vector bosons up to the first 12 KK excitations for $WW$ 
and ${\bar W} {\bar W}$ scattering
processes to examine the unitarity violation. The parameters $U_{E^4}^W$
and $U_{E^2}^W$ as well as $U_{E^4}^{\bar W}$
and $U_{E^2}^{\bar W}$ are listed in Table \ref{tableuv}.

\begin{table}[ht]
\begin{center}
\begin{tabular}{| c| c| c| c| c| c| c| c| c| c| c| c| c|}\hline
& & & & & & & & & & & & \\
  &$g_\gamma$ &$g_Z$ &$g_{\bar Z}$ & $g_{\gamma^\prime}$ & $g_{Z^\prime}$& $g_{{\bar Z}^\prime}$& $g_{\gamma^{\prime\prime}}$ & $g_{Z^{\prime\prime}}$ & $g_{{\bar Z}^{\prime\prime}}$& $g_{\gamma^{\prime\prime\prime}}$ & $g_{Z^{\prime\prime\prime}}$ & $g_{{\bar Z}^{\prime\prime\prime}}$\\
& & & & & & & & & & & &\\ \hline
& & & & & & & & &  & & &\\
$P_1$ &$0.3124$ &$0.6042$ &$-0.0364$ & $-0.0281$ & $0.0000$ & $-0.0014$ & $-0.0010$ & $-0.0000$ & $-0.0002$& $-0.0002$ & $0.0000$ & $0.0000$  \\
& & & & & & & & & & & &\\ \hline \hline
& & & & & & & & & & & &\\
$P_2$ &$0.3124$ &$0.6036$  &$-0.0775$  & $-0.0479$  &  $0.0040$& $0.0072$& $0.0060$ & $0.0000$ & $0.0008$& $0.0006$ & $0.0000$ & $0.0000$ \\
& & & & & & & & & & & &\\ \hline
& & & & & & & & & & & &\\
$P_3$ &$0.3124$ &$ 0.5940$ &$-0.0111$ & $-0.0526$ & $-0.0057$ & $0.0015$ & $0.0066$ & $-0.0038$ & $0.0002$ & $0.0002$ & $-0.0004$ & $0.0000$ \\
& & & & & & & & & & & &\\ \hline \hline
& & & & & & & & & & & &\\
$P_4$ &$0.3124$ &$0.5802$ &$-0.0109$ & $-0.0094$ & $-0.0014$ & $-0.0037$ & $-0.0032$ & $-0.0003$ & $-0.0017$& $-0.0015$ & $-0.0001$ & $-0.0008$ \\
& & & & & & & & & & & &\\ \hline
& & & & & & & & & & & &\\
$P_5$ &$0.3124$ &$0.5862$ &$-0.0032$ & $-0.0222$ & $-0.0109$ & $0.0003$ & $-0.0004$ & $-0.0052$ & $0.0000$& $-0.0014$ & $-0.0009$ & $ -0.0001$ \\
& & & & & & & & & & & &\\ \hline 
\end{tabular}
\end{center}
\caption{ Triple gauge couplings of $NWW$, up to the first 12 neutral vector bosons.}
\label{table3pw}
\end{table}

\begin{table}[ht]
\begin{center}
\begin{tabular}{| c| c| c| c| c| c| c| c| c| c| c| c| c|}\hline
& & & & & & & & & & & & \\
&${\bar g_\gamma}$ &${\bar g_Z}$ &${\bar g_{\bar Z}}$ &  ${\bar g_{\gamma^\prime}}$ & ${\bar g_{Z^\prime}}$& ${\bar g_{{\bar Z}^\prime}}$& ${\bar g_{\gamma^{\prime\prime}}}$& ${\bar g_{Z^{\prime\prime}}}$ & ${\bar g_{{\bar Z}^{\prime\prime}}}$ & ${\bar g_{\gamma^{\prime\prime \prime}}}$& ${\bar g_{Z^{\prime\prime\prime}}}$ & ${\bar g_{{\bar Z}^{\prime\prime\prime}}}$\\
& & & & & & & & & & & & \\ \hline
& & & & & & & & & & & & \\
$P_1$ &$0.3124$ & $0.2229$ & $1.3360$ & $1.1144$& $-0.0925$& $0.4284$ &$0.3416$ &  $-0.0170$ & $0.0072$ & $0.0033$  & $0.0007$ & $0.0001$ \\
& & & & & & & & & & & & \\ \hline \hline
& & & & & & & & & & & & \\
$P_2$ &$0.3124$ & $0.2969$ &$0.7910$  & $0.6349$& $-0.0416$ & $0.1836$ & $0.1431$ & $0.0002$ & $0.0055$& $0.0036$  & $0.0014$ & $-0.0018$ \\
& & & & & & & & & & & & \\ \hline
& & & & & & & & & & & & \\ 
$P_3$ &$0.3124$ & $0.2363$ & $0.1350$ & $1.7072$& $-0.0346$& $0.0954$ & $0.5356$ & $0.0060$ & $0.0035$ & $0.0069$  & $0.0054$ & $-0.0004$ \\
& & & & & & & & & & & & \\ \hline \hline
& & & & & & & & &  & & & \\
$P_4$ &$0.3124$ & $0.1028$ & $1.5638$ & $1.3228$& $-0.4006$& $0.5832$ & $0.4711$ & $-0.1663$ & $-0.0288$ & $-0.0309$  & $-0.0004$ & $-0.0006$ \\
& & & & & & & & & & & & \\ \hline
& & & & & & & & &  & & & \\
$P_5$ &$0.3124$ & $0.1625$ & $0.2171$ & $2.0184$& $-0.2039$& $0.0963$ & $0.7591$ & $-0.0257$ & $-0.0012$ & $-0.0376$  & $0.0277$ & $-0.0016$ \\
& & & & & & & & & & & & \\ \hline 
\end{tabular}
\end{center}
\caption{ Triple gauge couplings of $N {\bar W} {\bar W}$, up to the first 12 neutral vector bosons.}
\label{table3pwb}
\end{table}

\begin{table}[ht]
\begin{center}
\begin{tabular}{|  c| c| c| c| c| c| c| c| c|}\hline
 & & & & & & & & \\
  &$g_{Z W {\bar W}}$ &$g_{4W}^2$ & $g_{4{\bar W}}^2$ & $g_{ZZWW}^2$ & $U_{E^4}^W$ & $U_{E^2}^W$ &$U_{E^4}^{\bar W}$ & $U_{E^2}^{\bar W}$\\
& & & & & & & & \\ \hline
& & & & & & & & \\
$P_1$ &$-0.0532$  &$0.4648$ & $3.4834$& $ 0.3679$ & $0.0000$ & $0.0000$ & $0.0000$ & $0.0000$ \\
& & & & & & & & \\ \hline \hline
& & & & & & & & \\
$P_2$ &$-0.1094$ &$0.4704$ & $1.2705$ & $0.3765$ & $0.0000$ & $0.0000$ & $0.0000$ & $0.0004$ \\
& & & & & & & & \\ \hline
& & & & & & & & \\
$P_3$ &$-0.0626$ &$0.4537$ & $3.3847$ & $0.3572$ & $0.0003$ & $0.0623$ & $0.0011$ & $0.0210$ \\
& & & & & & & &\\ \hline \hline
& & & & & & & &\\
$P_4$ &$-0.0165$ &$0.4345$ & $5.0552$ & $0.3370$ &$0.0000$ & $0.0265$ & $0.0000$ & $0.0015$ \\
& & & & & & & &\\ \hline
& & & & & & & &\\
$P_5$ &$-0.0257$ &$0.4422$ & $4.9050$ & $0.3449$ & $0.0003$ & $0.1335$ & $0.0300$ & $0.2275$ \\
& & & & & & & &\\ \hline 
\end{tabular}
\end{center}
\caption{ Part of TGC, QGC, and unitarity violation parameters.}
\label{tableuv}
\end{table}

We notice that unitarity is almost exact for both $WW$ and ${\bar W} {\bar W}$ 
scattering processes for the points 1, 2, and 4. For points 3 and 5, we need to sum more
contributions of KK excitations. Generally speaking, 
the $U_{E^4}$ is smaller than $U_{E^2}$ for both $WW$ and ${\bar W} {\bar W}$ when we sum the
same number of KK resonances.
Meanwhile, the $U_{E^4}$ for $WW$ cases are smaller than 
that for ${\bar W} {\bar W}$ cases.
The $U_{E^2}$ for $WW$ cases are smaller than 
that for ${\bar W} {\bar W}$ cases. 
The unitarity violation parameters show that unitarity is guaranteed 
by KK resonances quite well.


\begin{figure}[t]
\centerline{
\includegraphics[width=8cm]{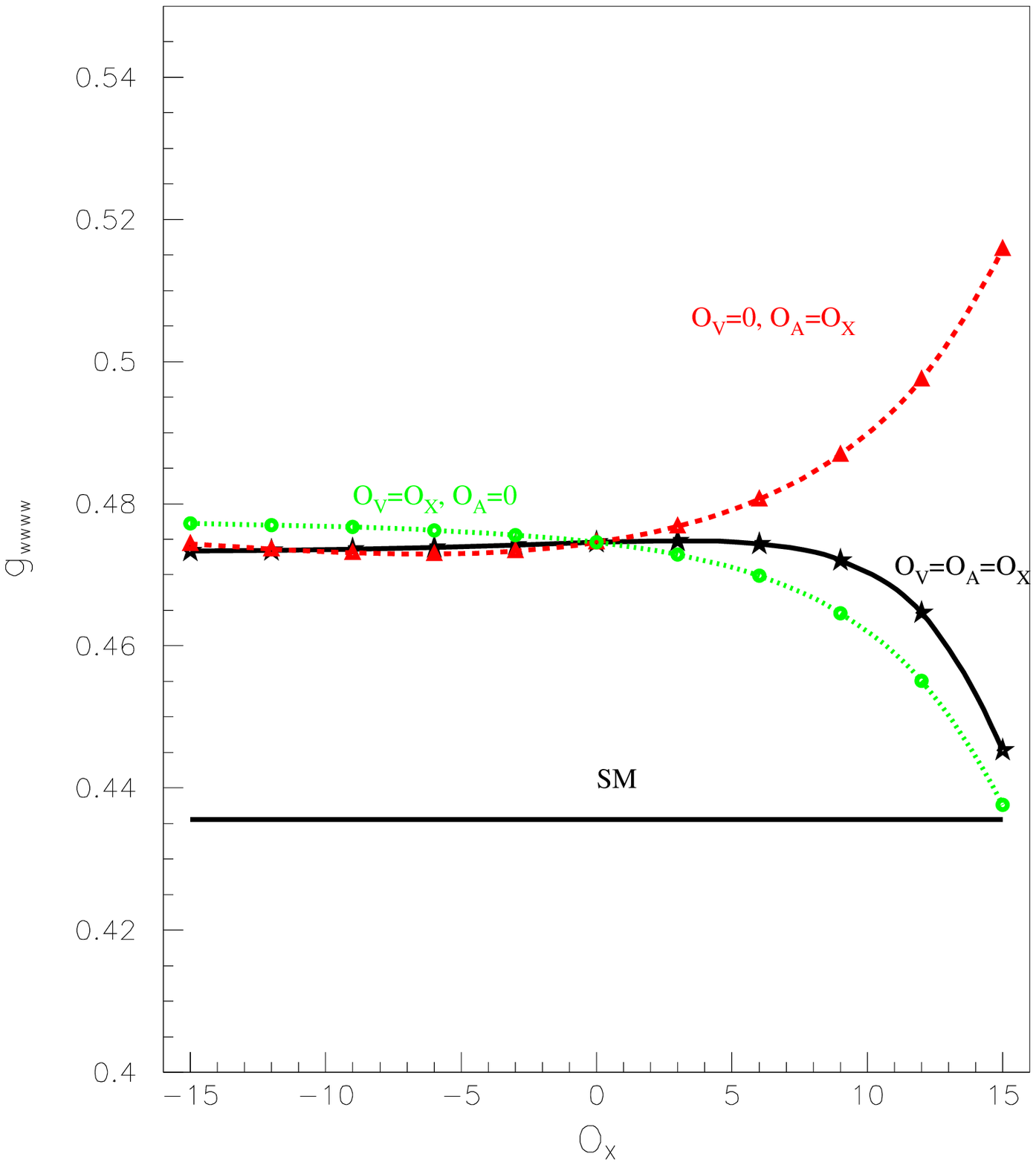}
\hspace*{0.0cm}
\includegraphics[width=8cm]{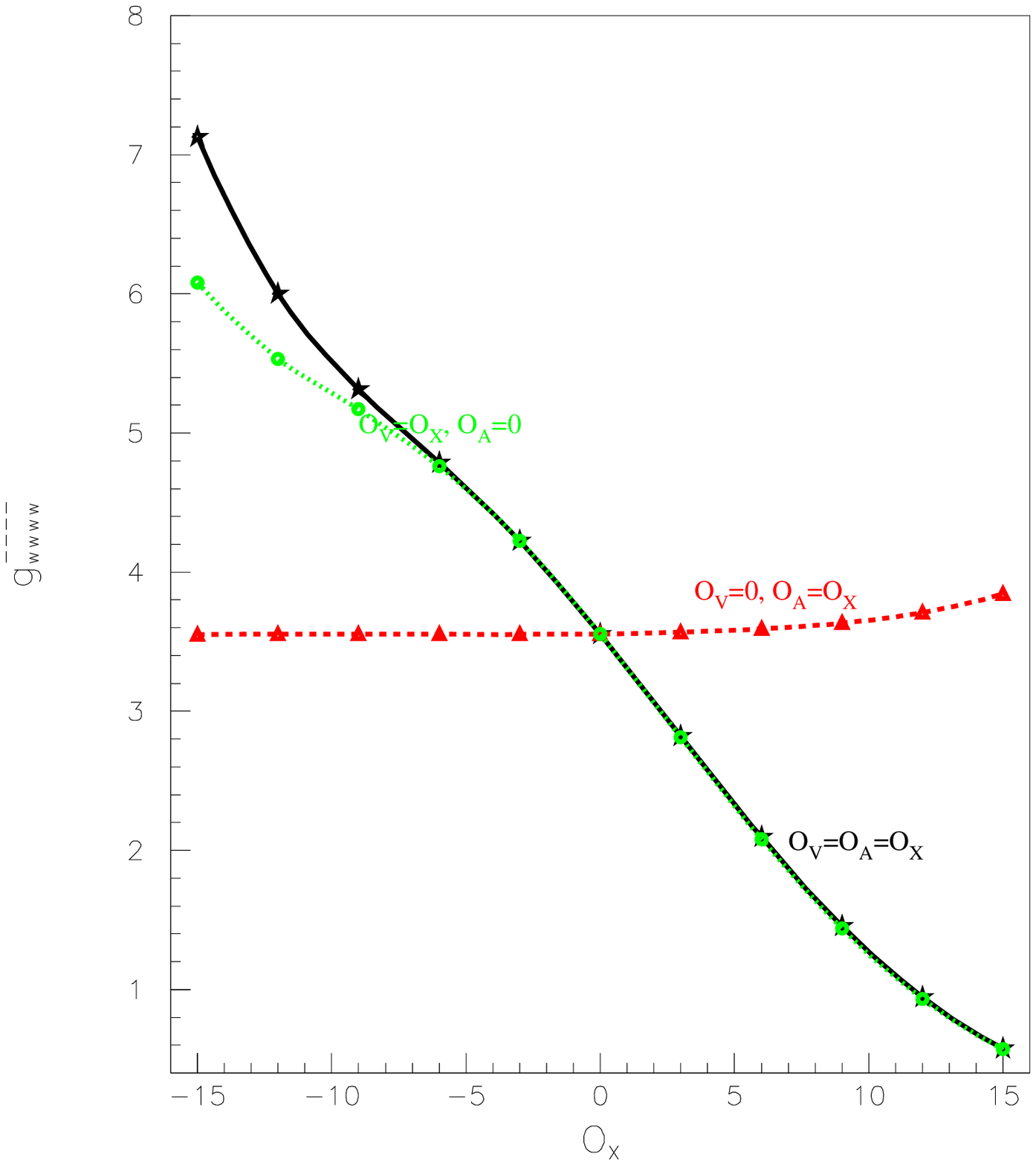}
}
\vskip -.25cm
\hskip 0.8 cm {\bf ( a) } \hskip 7.5cm {\bf (b)}
\caption{\it
Quartic gauge couplings, $g_{WWWW}^2$ and $g_{{\bar W}{\bar W}{\bar W}{\bar W}}^2$, 
vary with $O_X$. We select three cases to show the effects of $O_X$: 1) $O_V=O_A=O_X$,
2) $O_V=O_X, O_A=0$, and 3) $O_V=0, O_A=O_X$. Here the $r_b$  and $r_r$ are fixed
to $1$.
 }
\label{fig3}
\end{figure}

\section{Conclusion and Discussions}

In this paper, we parametrize the Higgsless models in the Hirn-Sanz scenario.
We study the effects of QCD improved background metric to
the spectra of KK resonances and find the spectra can be significantly modified.
We also study the three-point and four-point
couplings of vector bosons in the parameter space.
Roughly speaking, $ 10 \%$ deviations in TGC and QGC from the 
prediction of the SM can occur, which are within the detection capability of the 
LHC and ILC. Though it is hard to detect the KK excitations, the measurement of 
these TGC and QGC can provide complementary 
information on vector boson structure of the underlying theories.

On the new resonances sector, we notice that the couplings of $Z{\bar W}  W $ and $NWW$ vertices 
are usually smaller than the unitarity bounds. 
According to the simulation in Ref. \cite{Birkedal:2004au},
although the coupling of $Z{\bar W}  W $ is suppressed a little bit compared with
the unitary bounds, the total cross section for the production of ${\bar W}$ 
is large enough for its detection at the LHC.
According to the simulation of Ref. \cite{Malhotra:2006sx},
with the total integrated luminosity of $300$ fb$^{-1}$, 
it is possible to discover ${\bar Z}$ or ${\gamma'}$.
However, in the model we have studied here, 
the couplings of $NWW$ are suppressed at least $25\%$ for ${\bar Z}$ (
correspondingly the cross section will be suppressed by one order or so)
due to the almost degeneracy of ${\bar Z}$ or ${\gamma'}$. 
Therefore it might challenge the discovery of ${\bar Z}$ or ${\gamma'}$ via s-channel
at the LHC if ${\bar Z}$ or ${\gamma'}$ are larger than $500$ GeV. 
Recently, H.J. He {\it et al} \cite{He:2007ge} simulated the reconstuction
of $W'$ in a minimal Higgsless model at LHC with $P P \rightarrow W_0 Z_0
j j \rightarrow \nu 3 \ell jj$ scattering process
and found that there exists a very large
cancellation between fusion and non-fusion contributions, 
both of which are gauge dependent. Hence, the gauge-invariant 
results are substantially smaller than those 
given in the naive sum rule approach \cite{Birkedal:2004au, Malhotra:2006sx} 
in the unitary gauge. 
Nevertheless, the detection of new resonances, like $W'$, according to
the results of \cite{He:2007ge}, is still rather promising at LHC.

In the phenomenological models of Refs. \cite{Hirn:2006nt},
adding new operators in 5D Lagrangian seems arbitrary and higher 
order effects seems uncontrollable due to the lack of 
well-established power counting rules. Following the 
standard construction of the chiral effective Lagrangian, we should also introduce 
terms like $L_{MN} D^{M} D^{N}$ and 
$R_{MN} D^{M} D^{N}$ ( $D^M$ is the standard covariant 
differential operator defined in 5D ) in our consideration, 
which are counted as of the same order as the kinetic mixing term 
and can affect our results on TGCs and QGCs. Since the signs and magnitudes of these
effective couplings are unknown, the contributions of these operators can shift
our results in either directions.
However, in the present work, even without these operators, we observe large 
deviations from the predictions of the SM. 

In comparison, in the holographic models 
rooted in D-brane configurations of string theory, like the 
Sakai-Sugimoto model \cite{Sakai:2004cn,Sakai:2005yt}, the large-N and curvature 
corrections are well-defined and calculable. 
In the Sakai-Sugimoto model, the spontaneous symmetry breaking can be 
naturally realized with brane picture. Therefore it is interesting to extend 
our study to Sakai-Sugimoto model.

{\bf Notes:}
When this work is almost finished, we noticed that 
triple gauge couplings are also analyzed by \cite{Hirayama:2007hz}
in a technicolor model with Sakai-Sugimoto background metric.

\section{Acknowledgment}
We thank P. Ko for helpful discussion on the higher order operators
and careful reading of the manuscript, and H.J. He for communication
on ideal Fermion delocalization and the LHC sensitity to Higgless models.
The work of K.C. and Q.Y. was supported in part
by the NSC of Taiwan (NSC 95-2112-M-007-001
and 96-2628-M-007-002-MY3).

\end{document}